\documentclass[Physsubmission, Phys]{SciPost}

\hypersetup{
    colorlinks,
    linkcolor={red!50!black},
    citecolor={blue!50!black},
    urlcolor={blue!80!black}
}

\usepackage[bitstream-charter]{mathdesign}
\DeclareSymbolFont{usualmathcal}{OMS}{cmsy}{m}{n}
\DeclareSymbolFontAlphabet{\mathcal}{usualmathcal}

\begin{document}

\begin{center}{\Large \textbf{
Collective phenomena in pp interactions with high multiplicity \\
}}\end{center}

\begin{center}
N. Barlykov \textsuperscript{1,2},
V. Dudin \textsuperscript{1,2},
V. Dunin \textsuperscript{1} 
E. Kokoulina \textsuperscript{1,3$\star$}, 
A. Kutov \textsuperscript{4}, and
V. Nikitin \textsuperscript{1}
\end{center}

\begin{center}
{\bf 1} JINR
6 Joliot-Curie St
141980 Dubna, Moscow region, Russia
\\
{\bf 2} INP, 1 Ibragimova st, Almaty,
050032, Kazakhstan
\\
{\bf 3} GSTU 
Prospect Octiabria, 48,
246746 Gomel, Belarus
\\
{\bf 4}  IPM Komi SC UrD RAS
Kommunisticheskaja st., 24,
167000 Syktyvkar, Russia
* kokoulina@jinr.ru
\end{center}

\begin{center}
\today
\end{center}


\definecolor{palegray}{gray}{0.95}
\begin{center}
\colorbox{palegray}{
  \begin{tabular}{rr}
  \begin{minipage}{0.1\textwidth}
    \includegraphics[width=30mm]{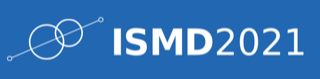}
  \end{minipage}
  &
  \begin{minipage}{0.75\textwidth}
    \begin{center}
    {\it 50th International Symposium on Multiparticle Dynamics}\\ {\it (ISMD2021)}\\
    {\it 12-16 July 2021} \\
    \doi{10.21468/SciPostPhysProc.?}\\
    \end{center}
  \end{minipage}
\end{tabular}
}
\end{center}
{\bf
The latest results on the search for collective phenomena in pp interactions at the U-70 accelerator at IHEP (Protvino) are presented. This the long-term experiment has been carried out at the SVD-2 setup.  We received the evidence of the pion (Bose-Einstein) condensate formation. Two noticeable peaks in the angular distribution of charged pions are interpreted as Cherenkov radiation of gluons by quarks. The leading effect disappears and the secondary system becomes isotropic in all directions at high multiplicity region.
}

\vspace{10pt}
\noindent\rule{\textwidth}{1pt}
\tableofcontents\thispagestyle{fancy}
\noindent\rule{\textwidth}{1pt}
\vspace{10pt}
\section{Introduction}
\label{sec:intro}
Multiparticle processes is still an actual direction in high energy physics. We are interested events with high multiplicity. These events are extremely rare. The average multiplicity at proton-proton collisions grows slowly as a logarithm of energy. 

Our experiment has been prepared and carried at the U-70 accelerator (IHEP in Protvino). It was aimed at studying of multiparticle production in the region close to the threshold production when almost all kinetic energy of colliding protons spend on the creation of secondaries. At the 50-GeV/c proton beam incident on a hydrogen target, mostly pions are produced, both charged and neutral. Their maximum possible number is about 59.

We have registered events with multiplicity significantly exceeding the average multiplicity (three or more times). These events are interesting in that one can find in them a manifestation of collective behaviour of secondaries. There are  quite a few theoretical predictions that indicate at such phenomena as the pionic (Bose-Einstein) condensate formation, Cherenkov radiation gluons, soft photon excess yield \cite{SP} and others phenomena.

\section{Searching for collective phenomena in high multiplicity events}
According to Gorenstein-Begun predictions \cite{Gor} the growth of the scaled variance, $\omega ^0 = D/\overline{ n_0}$ ($D$ is the variance of number of neutral pions, $\overline{ n_0}$ - their average multiplicity at given total multiplicity, $n_{tot}$) evidences the  pionic condensate formation. Our experimental data show a significant its growth beginning from $n_{tot } > $ 18 \cite{Jap}. 

\begin{figure}[h]
\centering
\includegraphics[width=0.3\textwidth]{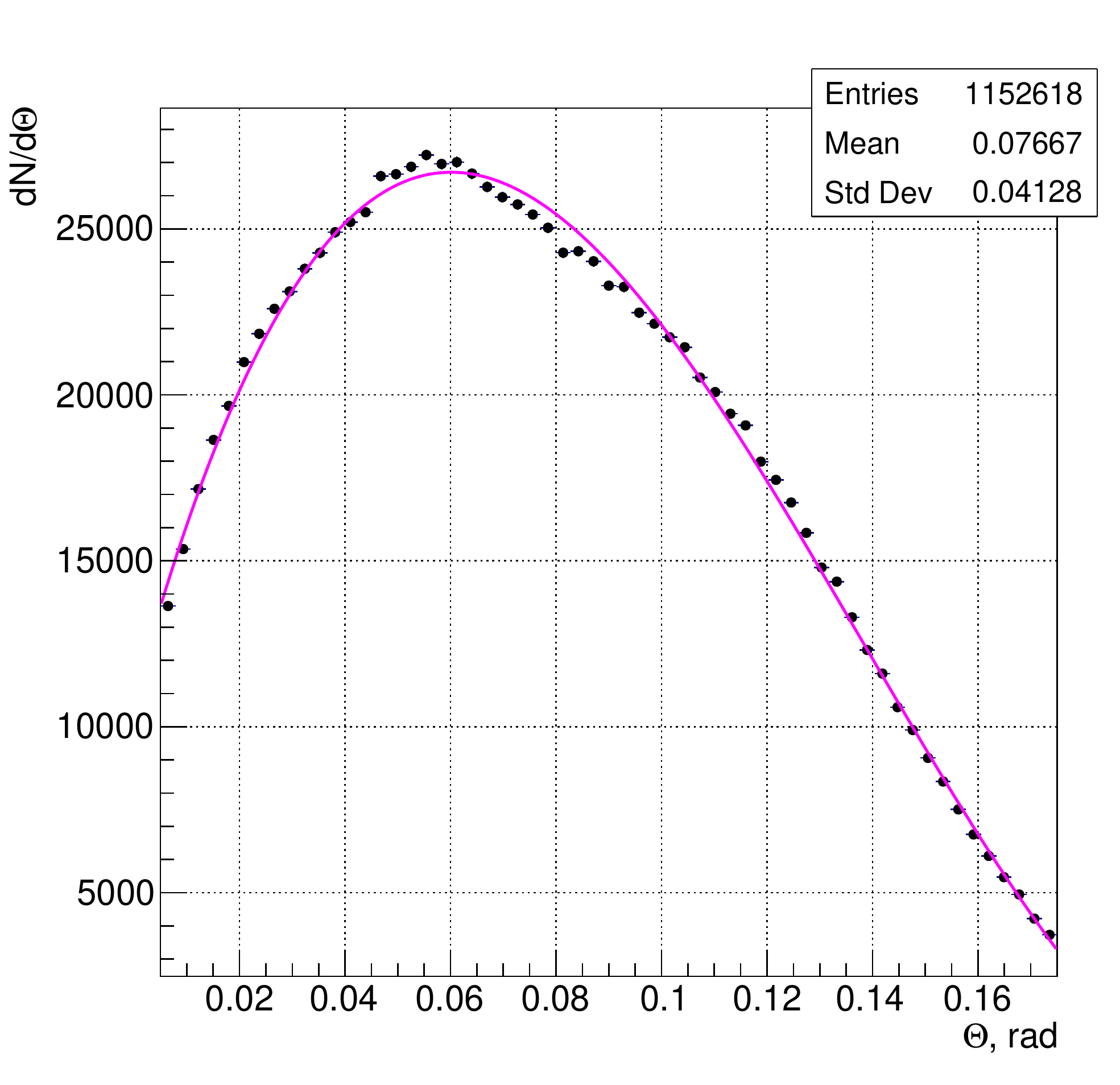}
\hspace{40pt}
\includegraphics[width=0.30\textwidth]{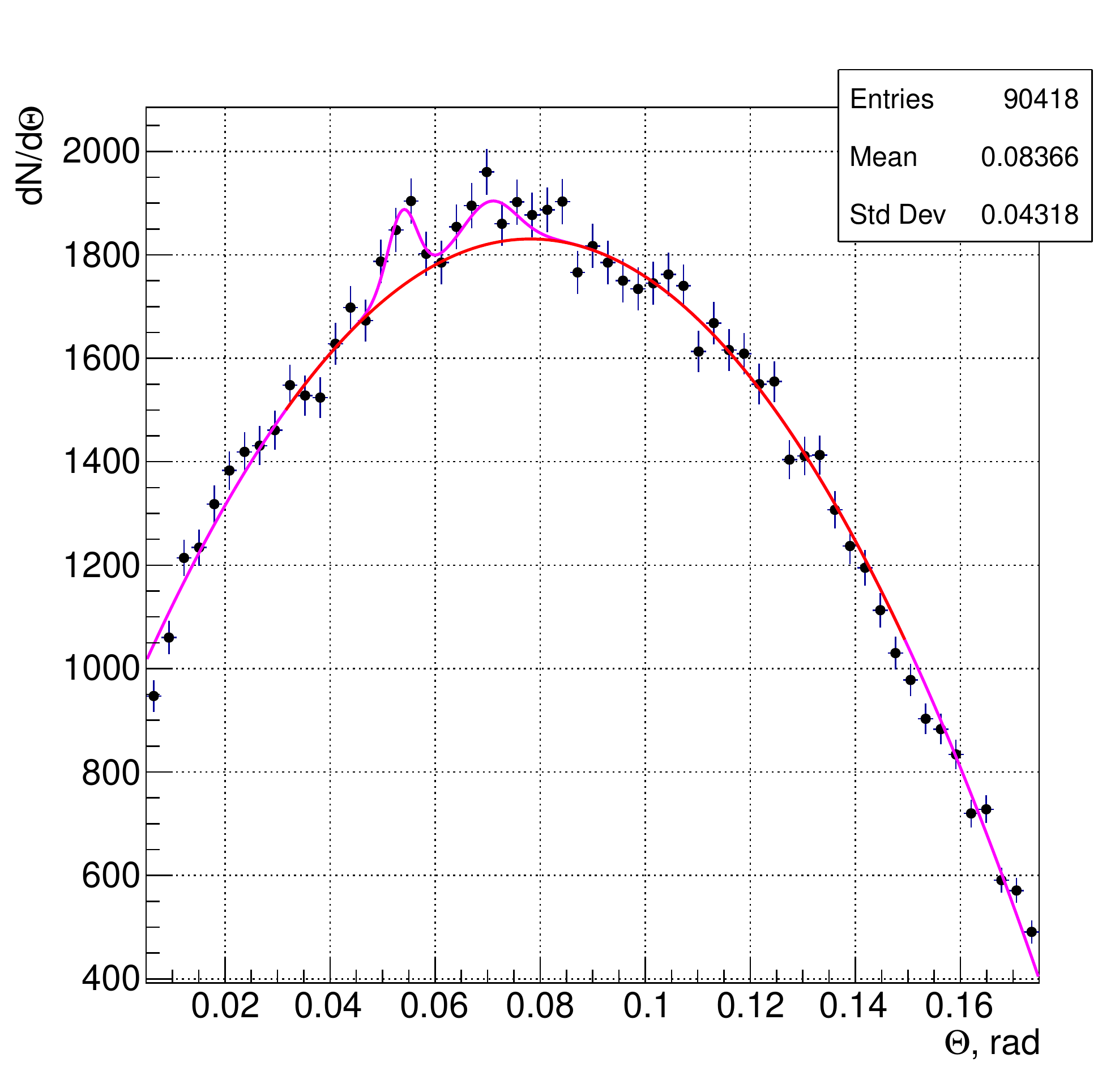}
\caption{Polar angle distribution for $n_{ch} $ $< $ 13 (left) and $n_{ch} $ $>$ 13 (right). Fit of two peaks is carried out by Gauss function, background -- by polynomial. 
}
\label{fig-1}
\end{figure}

The distributions  on the polar angle in two different region of multiplicity have shown that in the high multiplicity region we can observe the two-humped structure (Figure \ref {fig-1}, right) which we try to connect with Cherenkov radiation of gluons. That structure is absent in the low multiplicity region (Figure \ref {fig-1}, left) . 

Using Cherenkov's formula we estimate the refraction index of medium. Let $\Theta $ is the angle between primary and secondary tracks. The position of the left peak corresponds to  the polar angle $\Theta _{Cher }$ = 0.05377 $\pm $ 0.00273 rad with confidence 3.1 $\sigma $. For gluon rings, cos $\Theta $ = 1/$\beta n_r$ where $n_r$ is the refraction coefficient. For 50 GeV/c proton, we get $n_r$ = 1.0016 $\pm  $ 0.0001(4) for the left peak, i.e.  nuclear medium is highly rarefied. 

With growing of multiplicity the longitudinal component of momentum is decreasing, and the transversal one is staying the same. We expect that at the certain value of multiplicity both components will become equal. We define the region of multiplicity more than this value as a critical region at which both components of momentum (transversal and longitudinal) become indistinguishable. 

\begin{figure}[h]
\centering
\hspace{20pt}
\includegraphics[width=0.25\textwidth]{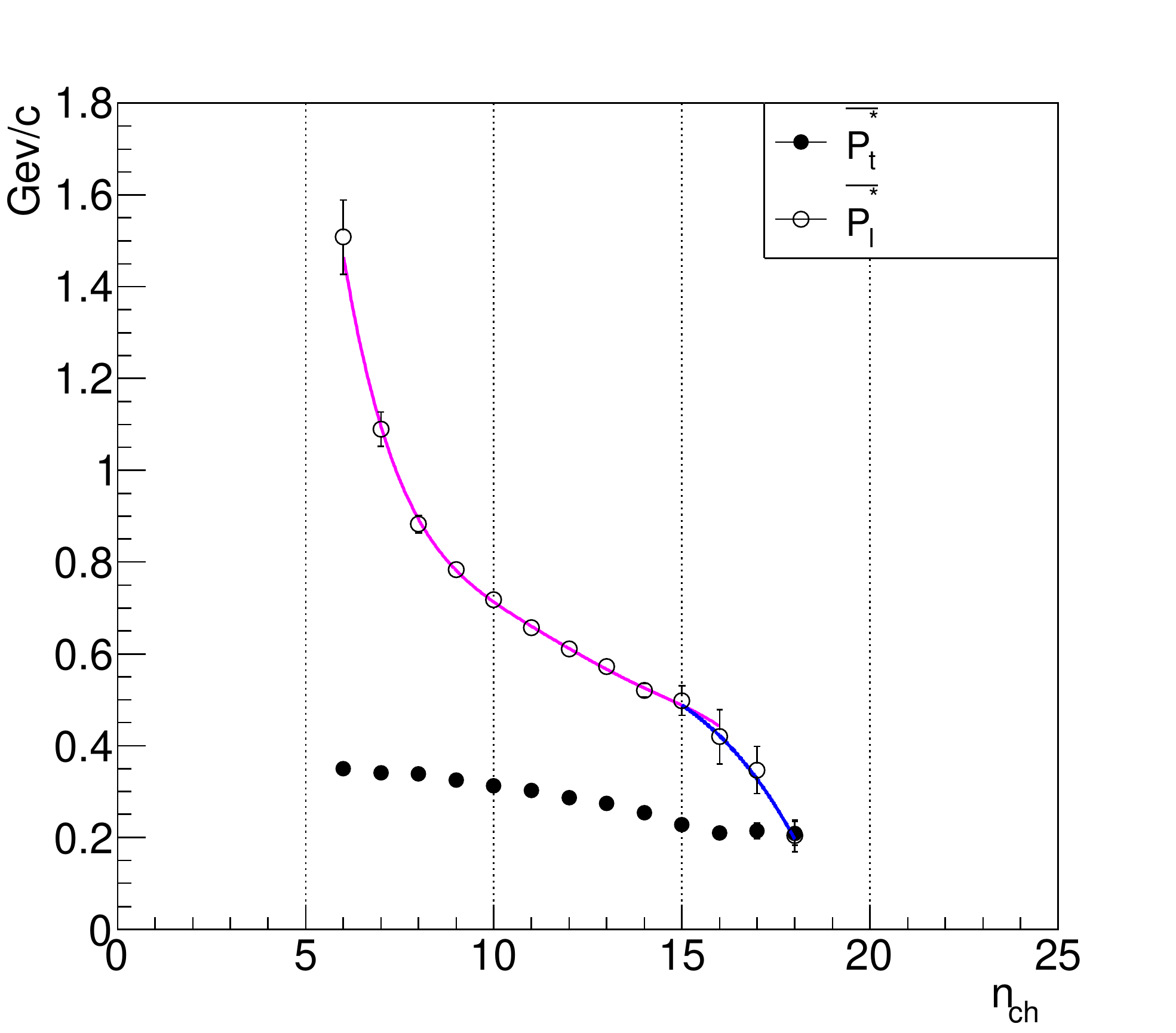}
\hspace{25pt}
\includegraphics[width=0.25\textwidth]{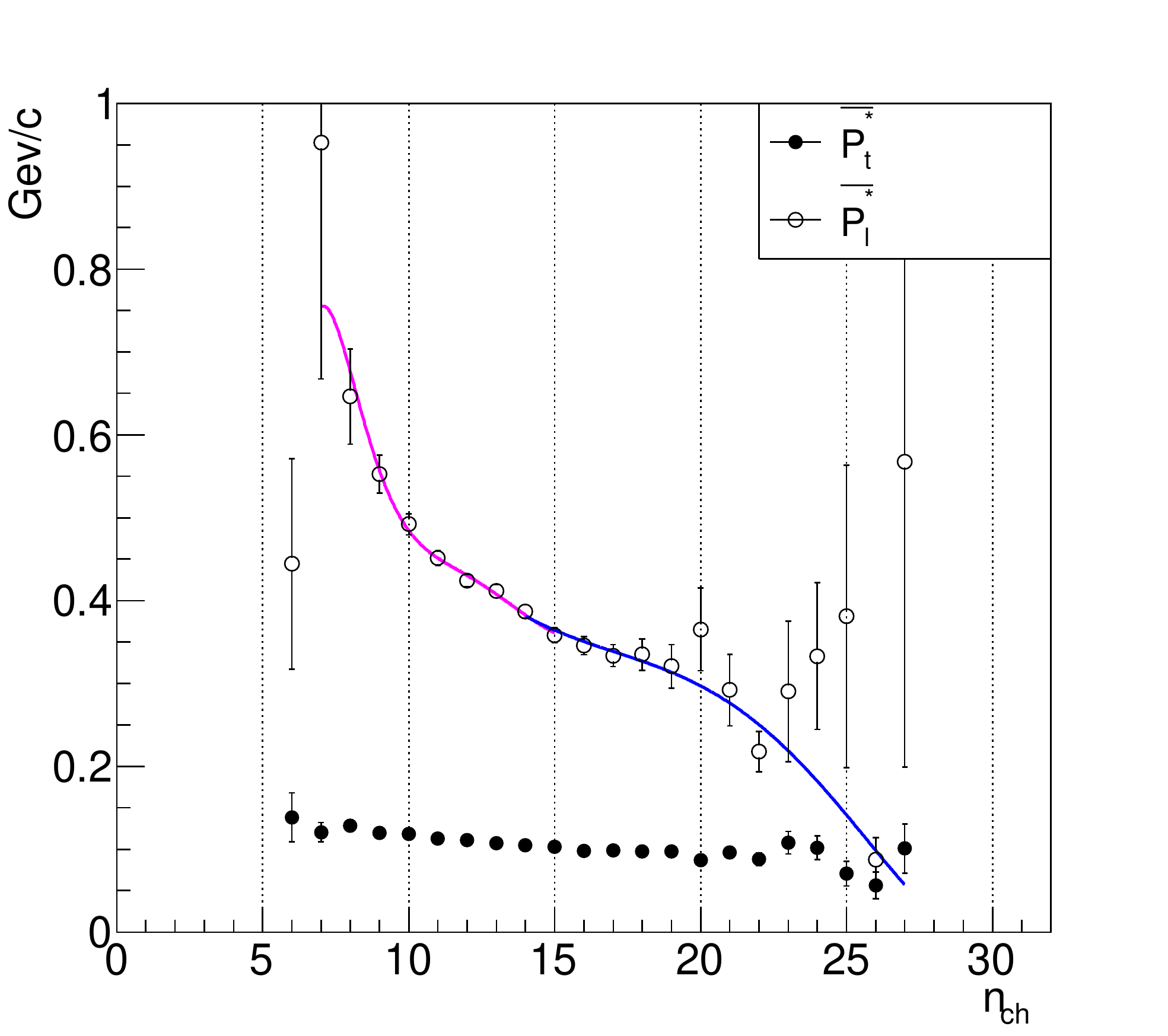}
\caption{(Left) The average values of longitudinal and transverse components of momentum of charged particles versus multiplicity for simulative events by PYTHIA 8. Red line describes them for $n_{ch}$ $<$ 14 (14 -- crossing of components), blue line -- more than 14.
(Right) The same components for experimental data.}
\label{fig-2}
\end{figure}

The average values of longitudinal ($\overline p^*_l$) and transverse ($\overline p^*_t$) components of momentum of charged particles in the center-of-mass system versus multiplicity $n_{ch}$ for simulative events by PYTHIA 8 are shown in Figure \ref {fig-2}, left. Red and blue lines describe longitudinal component for $n_{ch}$ $<$ 16  and $n_{ch}$ $>$ 16, respectively. The same dependences for experimental data (run 2008) with multiplicity corrections are presented in Figure \ref {fig-2}, right. Red and blue lines crossing corresponds to the critical multiplicity $n_{ch}$ = 14 (2 protons and 12 pions). 

Approaching of the longitudinal component to the transverse one can evidence the disappearance of the leading particle effect. This region of critical multiplicity found for charged particles corresponds to the region of total multilicity $n_{tot}$ $>$ 18 where pions fall down into the pion condensate.

We designed the gluon dominance model (GDM) \cite{GDM}. It presents convolution of a QCD branching gluons and their following hadronization. Gluon parameters in pp increase as opposed to e+e- annihilation that  is the evidence of the implementation of the recombination mechanism of hadronization in qg-medium but not in vacuum. The main sources of secondaries are active gluons and their branching leads to events with high multiplicity

\section{Conclusion}
High multiplicity study of pp interactions at U-70 have confirmed the existence of a series of collective phenomena. We evidence the Bose-Einstein (pionic) condensate formation in this region. The critical region of multiplicity where both momentum components are approaching to each other and the region of pionic condensate formation start from the same value of total multiplicity. 
The existence of the two-hump structure in the angular distribution of high multiplicity events can be interpreted as the gluon radiation in the rarefied parton medium
with the refraction index close to 1. The prepared experimental program of SP study at future accelerator complex, NICA, permits us to continue searching for collective phenomena in the high multiplicity region \cite{SP}

\end{document}